\documentclass[superscriptaddress,amsmath, amssymb, amsfonts, twocolumn, footinbib]{revtex4-1}
\usepackage[german,english]{babel}
\usepackage{graphicx}
\usepackage{graphics}
\usepackage{dcolumn}
\usepackage{bm}
\usepackage{amssymb}
\usepackage{amsmath}
\usepackage{amsfonts}
\usepackage{epsfig}
\usepackage{mathbbol}
\usepackage{latexsym}
\usepackage{dcolumn}
\usepackage{subfigure}
\usepackage{hyperref}
\usepackage{float}
\usepackage[normalem]{ulem}
\usepackage[usenames, dvipsnames]{color}
\usepackage{epstopdf}
\usepackage{comment}
\hypersetup{                    
    colorlinks,
    citecolor=red,
    filecolor=red,
    linkcolor=red,
    urlcolor=red
}

\newcommand {\pd} [2] {\frac{\partial #1}{\partial #2}}

 \newcommand {\beq}{\begin{equation}}
\newcommand {\eeq}{\end{equation}}
\newcommand {\beqn}{\begin{eqnarray}}
\newcommand {\eeqn}{\end{eqnarray}}
\newcommand {\bit}{\begin{itemize}}
\newcommand {\eit}{\end{itemize}}

\newcommand{\ba}{\begin{array}{rl}}
\newcommand{\ea}{\end{array}}
\newcommand{\bc}{\begin{cases}}
\newcommand{\ec}{\end{cases}}

\newcommand{\dps}{\displaystyle}

\newcommand{\om}{\iffalse}

\definecolor{mygray}{gray}{0.6}
\definecolor{gold}{RGB}{150, 150, 10}
\definecolor{mygreen}{RGB}{40, 200, 100}

\begin{document}
\renewcommand{\vec}{\mathbf}
\renewcommand{\Re}{\mathop{\mathrm{Re}}\nolimits}
\renewcommand{\Im}{\mathop{\mathrm{Im}}\nolimits}
\title{Anomalous drag in electron-hole condensates with granulated order}

\author{Hong Liu$^{1,2}$, Allan H. MacDonald $^{3}$  and Dmitry K. Efimkin$^{1,2}$}
\affiliation{$^1$School of Physics and Astronomy, Monash University, Victoria 3800, Australia}
\affiliation{$^2$ARC Centre of Excellence in Future Low-Energy Electronics Technologies, Monash University, Victoria 3800, Australia}
\affiliation{$^3$Center for Complex Quantum Systems, University of Texas at Austin, Austin, Texas 78712-1192, USA}

\begin{abstract}
We explain the strong interlayer drag resistance observed at low temperatures in bilayer electron-hole systems
in terms of an interplay between local electron-hole-pair condensation and disorder-induced 
carrier density variations.  Smooth disorder drive the condensate into a granulated phase in which interlayer coherence 
is established only in well separated and disconnected regions, or grains, within witch the densities of electrons and holes accidentally match. 
The drag resistance is then dominated by Andreev-like scattering of charge carries between layers at the grains 
that transfers momentum between layers. We show that this scenario can account for the observed dependence of the 
drag resistivity on temperature, and on the average charge imbalance between layers.
\end{abstract}
\date{\today}
\maketitle
\textcolor{blue}{\emph{Introduction}--} Thanks to progress in isolating and processing two dimensional 
materials, recent experiments~\cite{Tutuc-PRL-2018-E-H,Dmitry-PRB-2020,BECExc} have uncovered 
evidence of equilibrium condensation of spatially separated electrons (e) and holes (h) in the absence of a magnetic field, 
a phenomena first proposed some time ago~\cite{LozovikYudson1,Shevchenko}.
The zero-field electron-hole pair condensate state in semiconductor bilayers
is closely related to the charge density-wave states~\cite{CDWRecent,CDWKK,CDWMott,CDWRecentTBG}
of three-dimensional crystals, and to the electron-hole pair condensates
that occur for two-dimensional electrons in the strong magnetic field quantum Hall regime~\cite{EisensteinMacDonald,EisensteinReview}, but is expected to exhibit phenomenology that is distinct with both.  
Refs.~[\onlinecite{Tutuc-PRL-2018-E-H,Dmitry-PRB-2020}] reported strong enhancement of interlayer tunneling 
in a double bilayer graphene~\cite{Tutuc-PRL-2018-E-H,Dmitry-PRB-2020} and $\hbox{MoSe}_2$-$\hbox{WSe}_2$ heterostructure~\cite{BECExc}.
These observations on their own, however, demonstrate only local e-h coherence. 
Disorder is known to be deleterious for condensation, and it is not yet clear whether or not the quasi-long-range 
coherence and dipolar superfluidity that would potentially be useful for applications has been achieved.

In this Letter we address the influence of smooth disorder on the enhanced Coulomb drag~\cite{DragExp1,DragExp2,DragExp3,DragExp4, Tse2007,Hwang2011, MCarrega2012,Sch_drag_prl_2013,DragTheoryLevchenko, Hong-Drag-PRB-2017, 1999_Hall_drag,CDreview}  signal often used to detect electron-hole condensation~\cite{MacDonaldVignaleDrag}.  The drag resistance is defined as the ratio of the voltage drop that accumulates along an 
open layer to the current  driven through an adjacent layer.  When the bilayers are weakly coupled Fermi liquids, 
the drag resistance  has a quadratic temperature dependence 
at low temperatures~\cite{1977_drag, ee_drag_93_prb, Zheng-2DES-1993,1995_Hall_drag, DragFlensberg}.
Drag resistance $\rho_\mathrm{D}$ was predicted~\cite{MacDonaldVignaleDrag} to be colossally enhanced in the presence of 
a uniform e-h condensate and to experience a jump at the temperature of the Berizinskii-Kosterlitz-Thouless transition to the superfluid state~\cite{Berezinskii1,Berezinskii2,KosterlitzThouless1,KosterlitzThouless2}.  

\begin{figure}[b]
\vspace{-0.1 in}
\begin{center}
\includegraphics[trim=0cm 5cm 0cm 4cm, clip, width=1.0\columnwidth]{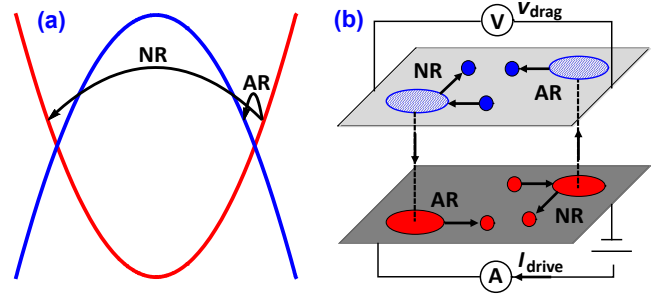}
\caption{\label{Two-layers} (a) Intralayer normal (NR) and interlayer Andreev-like (AR) reflections
of conduction band (red) and valance band (blue) electron states. 
AR involves interlayer tunneling and mediates momentum transfer between layers. (b) A drag device with  the driving current $I_\mathrm{drive}$ and  the drag voltage $V_\mathrm{drag}$. Inan  e-h condensate with granulated order
(grains are highlighted by red and blue regions), the Coulomb drag effect is dominated by AR. Due to the sensitivity of condensation to the local e-h imbalance, density variations across grains are minor compared to average e and h densities.  }
\end{center}
\vspace{-0.2in}
\end{figure}

Our model is motivated by experiments in conventional semiconductor quantum well (QW) and coupled graphene/QW systems~\cite{Christian-Lilly-PRB-2009,Seamons-Lilly-PRL-2009,Pepper-PRL-2008-drag,Pepper-PRB-2009-drag,Gamucci-Pellegrini-Natcomm-2014}.
These experiments  exhibit Coulomb drag signatures inconsistent with the Fermi liquid state scenario and are 
therefore indicative of strong e-h correlations.
In these experiments, $\rho_\mathrm{D}$ reaches a minimum as temperature is decreased, that is   
followed by growth and finally saturation at even lower temperatures. 
The observed upturn is much smaller than that predicted~\cite{MacDonaldVignaleDrag} in the case of 
uniform e-h condensate, and does not exhibit the strong sensitivity to the mismatch of e and h densities, which is the hallmark of 
BCS-like electron-hole pair condensation~\cite{EfimkinLOFF,SeradjehLOFF,MacDonaldMismatch}. 
The observed anomalous behavior has been variously interpreted in terms of fluctuating Cooper pairs, which are a precursor of e-h condensation~\cite{Hu,Mink1,Mink2,Dmitry-PRB-2013}, and as evidence for 
the formation of a nondegenerate  gas of interlayer excitons~\cite{Dmitry-PRL-2016}. These scenarios can qualitatively explain only some aspects of the experimental data, and the observed anomalous behavior is still far from understood.

Here we explain the anomalous drag effect in terms of an 
interplay between electron-hole condensation and large-spatial-scale density variations. The latter are common in two dimensional systems~\cite{EHpuddles,Puddles1,Puddles2,Puddles3}, but their importance for electron-hole coherence phenomena has not been emphasized previously.  We argue that carrier-density variations drive
the bilayer to a phase in which local condensation occurs in well separated disconnected patches with the densities of electrons and holes accidentally matching.
We demonstrate that the Coulomb drag effect in this state with \emph{granulated order}
is dominated by Andreev-like reflection of charge carries at the grains. Since the components of e-h Cooper-like pairs are spatially 
separated, the Andreev-like reflection process, illustrated in Fig.~1-a and -b, enables momentum transfer between layers. Fits of our theory to available experimental data demonstrates that this scenario can account for the dependence of 
$\rho_\mathrm{D}$  on temperature and e-h imbalance consistently.

\textcolor{blue}{\emph{The model}--} 
Recently, several researchers have developed microscopic 
theories~\cite{GrapheAfter1,GrapheAfter2,GrapheAfter3,GrapheAfter4,GaAsRecent,superfluidity-PRB-2013-Das,EfimkinEH}
and attempted to quantitatively predict the density/temperature phase diagrams of 
electron-hole bilayers in the absence of disorder. 
(See Ref.~\cite{EHreview} for a review.) 
Since the anomalous Coulomb drag effect has been observed 
in both semiconductor QW bilayers and in hybrid QW/graphene bilayers, electronic structure details do not play an essential role.
We therefore choose a phenomenological approach that is consistent with 
the experimentally relevant~\cite{GaAsRecent} weak-to-moderate coupling regime on the 
Bardeen-Cooper-Schrieffer side of the BCS-BEC crossover between weak-pairing and 
Bose-Einstein condensation of indirect excitons.

A bilayer with symmetric quadratic dispersion for $e$ (electrons) and $h$ (holes) is described by
\begin{equation}
\hat{H}_0=\int \vec{dr} \cdot  \left[ e^\dagger_{\vec{r}} (\xi^\mathrm{e}_\vec{p} + V_\mathrm{e}(\vec{r})) e_{\vec{r}} +  h^\dagger_{\vec{r}} (\xi^h_\vec{p} + V_\mathrm{h}(\vec{r})) h_{\vec{r}}\right].
\label{eq:h0}
\end{equation}
Here $e_\vec{r}$ and $h_\vec{r}$ are annihilation operators for electrons and holes, 
$\xi^\alpha_\vec{p}=\vec{p}^2/2m-\epsilon_\mathrm{F}^{\alpha}$ is the band dispersion, 
$m$ is the effective mass, and $\alpha=\mathrm{e}(\mathrm{h})$ is the layer index.
The Fermi energy $\epsilon_\mathrm{F}^\alpha$ determines the spatially average electron and hole densities, which can be varied using external gates.   
In Eq.~(\ref{eq:h0}) $V_\alpha(\vec{r})$ is the random potential responsible for the long-range density variations.  We take $\langle V_\alpha(\vec{r})\rangle=0$, and the disorder correlation function
$\langle V_{\alpha'}(\vec{r'})V_\alpha (\vec{r})\rangle = \delta_{\alpha',\alpha}\;
G_0(|\vec{r}-\vec{r}'|)$ with 
\begin{equation}
\label{GF}
G_0(\vec{r})=w^2\text{exp}\big[-r^2/\xi^2\big].
\end{equation}
Here $w$ is the average height of disorder potential $V_\alpha(\vec{r})$ and $\xi$ is its correlation length.
Our neglecting of disorder correlations between layers is likely to be physically realistic but would in any case only play
an important role if potentials in the two layers were nearly identical~\footnote{It also should be noted that the model with the density variations only in one layer is also analytically tractable, and the corresponding results can be obtained by the rescaling $w\rightarrow  w/\sqrt{2}$. This suggests that the possible asymmetry between density variations in two layers weakly affects main results and conclusions of this work.}.

The interlayer interactions in the system can be described by an effective contact 
potential $U$ as follows:~\footnote{We assume that the only role of the intralayer interactions is 
renormalization of spectrum for electrons and holes.}
\begin{equation}
\label{Eq:MF}
H_U=\int_{\vec{r}} U e^\dagger_\vec{r} h^\dagger_{\vec{r}'} h_{\vec{r}'}e_\vec{r} \rightarrow \int_\vec{r}\left[\Delta(\vec{r}) e^\dagger_\vec{r} h^\dagger_\vec{r}+h.c.\right]. 
\end{equation}
The second form for the right-hand-side of Eq.~(\ref{Eq:MF})
makes a mean-field approximation and introduces the complex electron-hole pair order 
parameter $\Delta(\vec{r})$.
Cooper pair condensation is known to be quite sensitive to a density mismatch between electrons and holes.
We assume that local pairing is strongly suppressed when the mismatch of the local 
Fermi energies for electrons and holes $\delta(\vec{r})=\delta_\mathrm{F}+\delta_\mathrm{V}(\vec{r})$ exceeds a phenomenological chosen temperature-dependent critical mismatch $\delta_0\ll\epsilon_\mathrm{F}$. Here we have separated the mismatch into a spatially averaged contribution
$\delta_\mathrm{F}=\epsilon_\mathrm{F}^\mathrm{e}-\epsilon_\mathrm{F}^\mathrm{h}$ and a spatially varying one $\delta_\mathrm{V}(\vec{r})=V_\mathrm{e}(\vec{r})-V_\mathrm{h}(\vec{r})$. The behavior of the system then depends on the relation between $\delta_0$, $w$ and $\epsilon_\mathrm{F}$. 

When $w\ll \delta_0$ the effect of density variations is minor and a uniform BCS-like state is favored. 
When $w\sim \delta_0$ we expect a complicated interplay between uniform-mismatch Larkin-Ovchinnikov-Fulde-Ferrell physics~\cite{LO,FF} and 
randomness due to density imbalance spatial variation. 
We focus on the case $\delta_0 \ll w\ll \epsilon_\mathrm{F}$,
in which density variations are still minor, but strongly impact e-h condensation, which survives only in well separated disconnected regions, or grains, where e and h densities accidentally match
~\footnote{The granulated behavior has also been predicted in non equilibrium Bose-Einstein condensates~\cite{Yukalov1,Yukalov1}, however its origin is completely different compared to the one for the granulated order in this Letter.}. In this regime anomalous Coulomb drag effect has been reported both in semiconductor QWs, and in hybrid graphene/QW bilayers. The regime of strong e-h imbalance variations $w\sim\epsilon_\mathrm{F}$ is realized in these systems at much lower charge carrier densities and corresponds to a percolation transition to the truly granulated insulating state~\cite{Single-QW-Lilly,PhysRevLett.94.136401,PhysRevLett.100.016805} or to a presence of e-h puddles~\cite{EHpuddles,Puddles1,Puddles2,Puddles3}.

We will assume that in the considered regime $\delta_0 \ll w\ll \epsilon_\mathrm{F}$ phase coherence is maintained within grains, and that 
the amplitude of the order parameter $\Delta_\mathrm{A}$ adjusts to the local imbalance $\delta(\vec{r})$ as follows
\begin{equation}
\label{DeltaModule}
\Delta_\mathrm{A}(\vec{r})=\Delta_0 \mathrm{exp}[-\delta^2(\vec{r})/\delta^2_0].
\end{equation} 
Here $\Delta_0$ is temperature-dependent order parameter value in the absence of the mismatch. Below we examine the Coulomb drag effect in the presence of e-h condensate with granulated order. 

\begin{figure}[b]
\vspace{-0.1 in}
\begin{center}
\includegraphics[width=10cm]{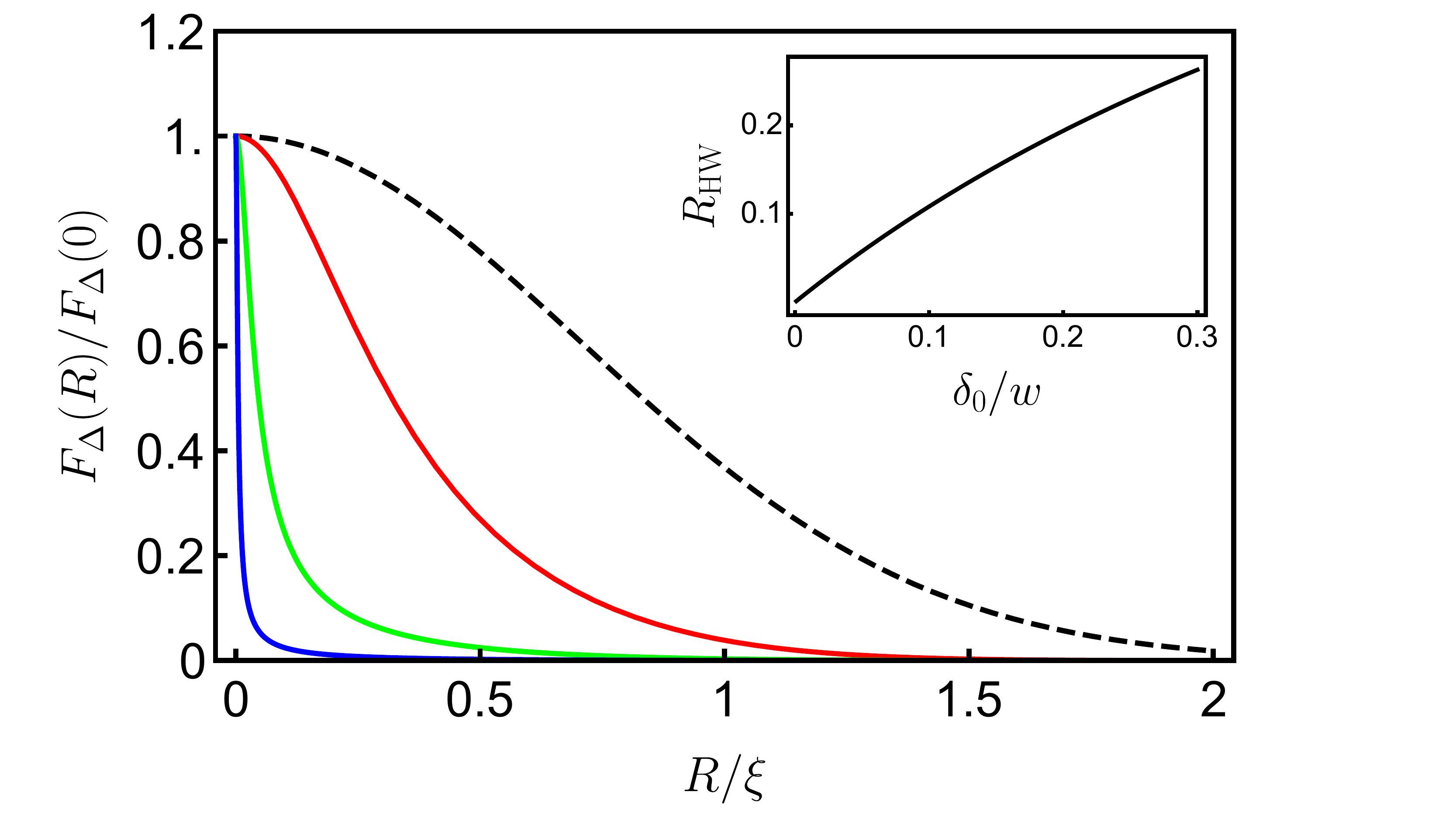}
\vspace{-0.2 in}
\caption{\label{Profile-0} The spatial profile of the correlation function $F_\Delta(\vec{R})/F_\Delta(0)$ in Eq.~(\ref{correlation}) in the balanced case $\delta_\mathrm{F}=0$ at $\delta_0/w=0.4$ (red), $0.04$ (green), and $0.004$ (blue). 
For $\delta_0\ll w$ its width $R_\mathrm{HW}$, defined at half-maximum and presented in the inset, 
is much smaller that the width $\xi$ of the disorder correlation function $G_0(R)/G_0(0)$ 
given by Eq.~(\ref{GF}) and plotted here as a dashed black line. This property helps to justify the picture of well separated e-h condensate grains. 
}
\end{center}
\vspace{-0.3 in}
\end{figure}

\textcolor{blue}{\emph{Andreev-like reflection from grains}--}
The granulated electron-hole condensate state 
does not support superfluidity, but its presence still can have a strong impact on drag since each grain can act as a local inter-layer scatterer via the Andreev-like reflection (AR) mechanism. Its interlayer nature is intricately connected with the spatial separation of the components of the Cooper-like pairs. As a result, AR mediates momentum transfer between layers and produces a drag 
effect that mimics Fermi liquid Coulomb drag but, as we now explain,
has a decidedly different temperature dependence.


To calculate the Andreev-like reflection probability, we treat the pair potential perturbatively and 
use a Fermi's golden rule for the scattering rate between momentum states in two layers,
\begin{equation}
P_{\vec{p} \vec{p}'}=\frac{2\pi}{\hbar} F_\Delta(\vec{p}-\vec{p}')\; \delta(\xi^\mathrm{h}_\vec{p}-\xi_{\vec{p}'}^\mathrm{e}),
\end{equation}
which is determined by the absolute value of the Fourier transform of the
condensate correlation function $F_\Delta(\vec{R})=\langle \Delta(\vec{R})\Delta^*(0)\rangle$. Within the picture of well separated grains 
that maintain local phase coherence, the scattering rate has independent contributions from each grain and 
no contribution from correlated scattering from different grains.
The scattering rate is determined only by the position dependent absolute value of the order 
parameter $\Delta_\mathrm{A}(\vec{r})$, and the correlation function can be rewritten as follows $F_\Delta(\vec{R})=\langle \Delta_\mathrm{A}(\vec{R}) \Delta_\mathrm{A}(0)\rangle- \langle \Delta_\mathrm{A}^2(0) \rangle$. If we use Eqs.~(\ref{DeltaModule}) and (\ref{GF}), the calculation of the correlation function reduces to the evaluation of a Gaussian integral.

In the absence of the $\mathrm{e}$-$\mathrm{h}$ imbalance ($\delta_\mathrm{F}=0$),
$F_\Delta(\vec{R})=\Delta_0^2 \exp[-\Omega(\vec{R})]$ 
where $\Omega(\vec{R})$ is an effective thermodynamic potential of a classical field $\delta_\mathrm{V}(\vec{r})$ 
interacting with two repulsive centers. The latter are described by the potential $P(\vec{r})=[\delta_\mathrm{D}(\vec{r})+\delta_\mathrm{D}(\vec{r}-\vec{R})]/\delta_0^2$ where $\delta_\mathrm{D}(\vec{r})$ is the Dirac delta-function. As the a result, the corresponding effective action is given by 
\begin{equation}
S=\int_\vec{r} P(\vec{r})\delta_\mathrm{V}^2(\vec{r}) + \frac{1}{2}\int_{\vec{r}\vec{r}'}\delta_\mathrm{V}(
\vec{r})G^{-1}_0({\vec{r}-\vec{r}'})\delta_\mathrm{V}(\vec{r}'). 
\end{equation}
A straightforward Gaussian integration results in 
\begin{equation}
\Omega(\vec{R})=\frac{1}{2}\mathrm{tr}[\ln(1-\hat{P} \hat{G_0} )]=\frac{1}{2}\int_0^1 d\lambda\; \mathrm{tr} [\hat{P} \hat{G}].
\end{equation}
Here we introduced an auxiliary coupling constant $\lambda$ 
by letting $P(\vec{r})\rightarrow \lambda P(\vec{r})$ and employed the 
coupling constant integration method. 
The latter allows us to reorder and re-sum the perturbation series in terms of a renormalized Green function $\hat{G}$ which satisfies
the Dyson equation $\hat{G}=\hat{G}_0 - \hat{G}_0 \lambda \hat{P} \hat{G}$. Since the potential $P(\vec{r})$ represents a sum of two point-like scattering centers the Dyson equation is algebraic and the calculation of $\Omega(\vec{R})$ is straightforward. This trick can be generalized to the presence of e-h imbalance  $\delta_\mathrm{F}$~\footnote{See Supplemental Material.}. As a result, the average amplitude of the order parameter $\langle\Delta_\mathrm{A} \rangle$ and the correlation function $F_\Delta(\vec{R})$ are given by
\begin{widetext}
\begin{equation}\label{correlation}
\langle\Delta_\mathrm{A} \rangle = \Delta_0 \frac{ \delta_0 e^{-\frac{\delta^2_\mathrm{F}}{\delta^2_0+2w^2}} }{\sqrt{\delta^2_0+2w^2}},  \quad \quad\quad  F_\Delta(\vec{R})=\Delta^2_0\Big\{\frac{ \delta^2_0 e^{-\frac{2\delta_\mathrm{F}^2}{\delta^2_0+2w^2(1+E)}}}{\sqrt{\big[\delta^2_0+2w^2(1+E)\big]\big[\delta^2_0+2w^2(1-E)\big]}}-\frac{\delta^2_0 e^{-\frac{2\delta^2_\mathrm{F}}{\delta^2_0+2w^2}}}{\delta^2_0+2 w^2}\Big\}.
\end{equation}
\end{widetext}
Here $E\equiv e^{-R^2/\xi^2}$.  The spatial dependence of the correlation function in the balanced case $\delta_\mathrm{F}=0$ is presented in Fig.~\ref{Profile-0}. In the considered regime $\delta_0\ll w$ it exponentially decays at the spatial scale $\xi_\mathrm{\Delta}=\sqrt{3} \delta_0 \xi/w$ which is much smaller than the spatial scale of variations $\xi$, justifying the picture of well separated e-h condensate grains.

Locally e-h pairing is very sensitive to the imbalance and is strongly suppressed if the latter exceeds $\delta_0$. However,  as it is clearly seen in Eq.~(\ref{correlation}), the \emph{average} order parameter $\Delta_\mathrm{A}$ and the correlation function $F_\mathrm{\Delta}$ are robust to the imbalance until it exceeds $w$. In this case the probability of finding a spot with matching charge carrier densities for
electrons and holes is exponentially small. 

\textcolor{blue}{\emph{Transport in the phase with granulated order}--} 
The transport properties of the bilayer can be described by coupled Boltzmann equations:  
\begin{equation}\label{Kinetic-E}
e {\vec{E}}_\alpha\cdot\pd{f^\alpha_{{\vec{p}}}}{\vec{p}}=-\frac{f_\alpha-\bar{f}_{\alpha}}{\tau_\alpha} + I_\mathrm{C}^\alpha[f^\mathrm{e},f^\mathrm{h}] + I_\mathrm{A}^\alpha[f^\mathrm{e},f^\mathrm{h}]. 
\end{equation}
Here $\vec{E}_\alpha$ is the electric field that disturbs the equilibrium distribution $\bar{f}^\alpha_\vec{p}$ of charge carriers, and $\tau_\alpha$ is the
transport relaxation time associated with short-range disorder within the layers, which we have so far disregarded. 
The distribution functions are coupled by interlayer Coulomb scattering and by Andreev-like scattering at the condensate grains.
We do not discuss $I_\mathrm{C}^\alpha[f^\mathrm{e},f^\mathrm{h}]$ since the resistivities produced by the 
two drag mechanisms have very different temperature dependence  and are approximately additive.
The Andreev-like scattering integral is 
\begin{equation}
I_\mathrm{A}^\alpha[f^\mathrm{e},f^\mathrm{h}]=\sum_{\vec{p}'}P_{\vec{p}\vec{ p}'}(f^\mathrm{e}_{\vec{p}}-f^\mathrm{h}_{\vec{p}'}).
\end{equation}\\
The contribution to $\rho_\mathrm{D}$ induced by AR at grains is
\begin{equation}
\label{RhoD}
    \rho^\mathrm{AR}_{\mathrm{D}}=\frac{\hbar}{e^2}\frac{1}{\sqrt{n_\mathrm{e} n_\mathrm{h}}}\frac{N^\mathrm{h}_\mathrm{F}\hbar}{\tau^\mathrm{h}_{\mathrm{AR}}}=\frac{\hbar}{e^2}\frac{1}{\sqrt{n_\mathrm{e}n_\mathrm{h}}}\frac{N^\mathrm{e}_\mathrm{F}\hbar }{\tau^\mathrm{e}_{\mathrm{AR}}},
\end{equation}
where $n_\alpha$ is the density for charge carriers and $N^\alpha_{\mathrm{F}}=m/2\pi \hbar^2$  is the corresponding density of states. 
The drag resistivity does not depend on the transport relaxation times $\tau_\alpha$ within each layer, but solely on the 
AR transport scattering time 
$\tau_\mathrm{tr}^\alpha$ for AR  at the Fermi level:
\begin{equation}
\frac{1}{\tau_\mathrm{AR}(\vec{p})}=\sum_{\vec{k'}} P_{\vec{p}\vec{p}'}\cos\phi_{\vec{p}\vec{p}'}.
\label{eq:artime}
\end{equation}
The $\cos$ factor in Eq.~(\ref{eq:artime}), is analogous to the $1-\cos(\phi_{\vec{p}\vec{p}'})$ factor that 
appears in the standard transport scattering time, and captures the fact that the signs of the contributions for 
forward and backward reflections are opposite. 

\begin{figure}[t]
\vspace{-0.1in}
\begin{center}
\includegraphics[width=7.5cm]{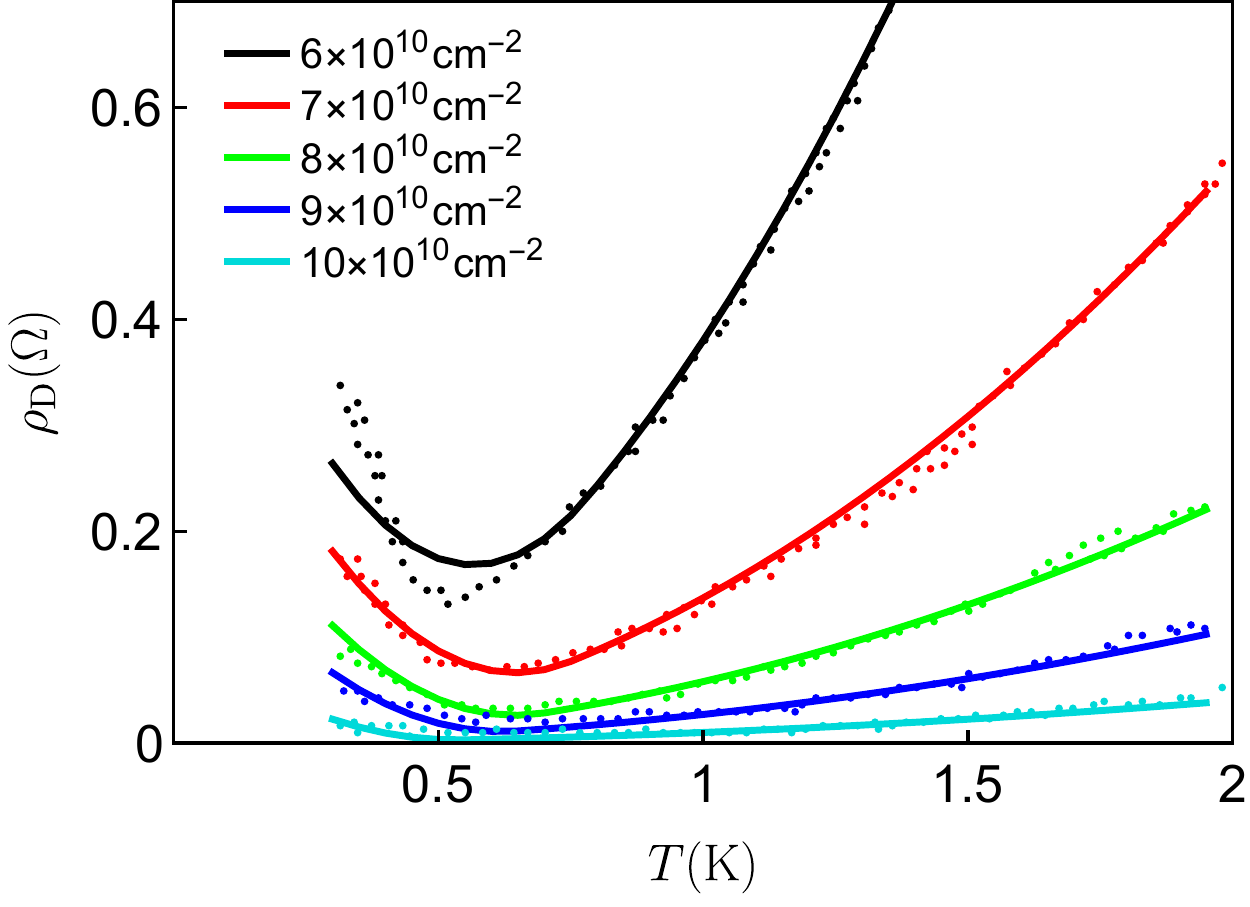}
\caption{\label{Balanced-density}The temperature dependence of the drag resistivity for equal densities of electrons and holes. The dotted lines are experimental data from Ref.~\cite{Seamons-Lilly-PRL-2009}. The real lines are fits to our model with disorder average height $w=1.9\mathrm{meV}$
and correlation length $\xi=35\mathrm{nm}$.  The $T_0$ local-pairing temperature is fitted separately for each curve.
}
\end{center}
\vspace{-0.4in}
\end{figure}

\begin{figure}[t]
\vspace{-0.0in}
\begin{center}
\includegraphics[width=8cm]{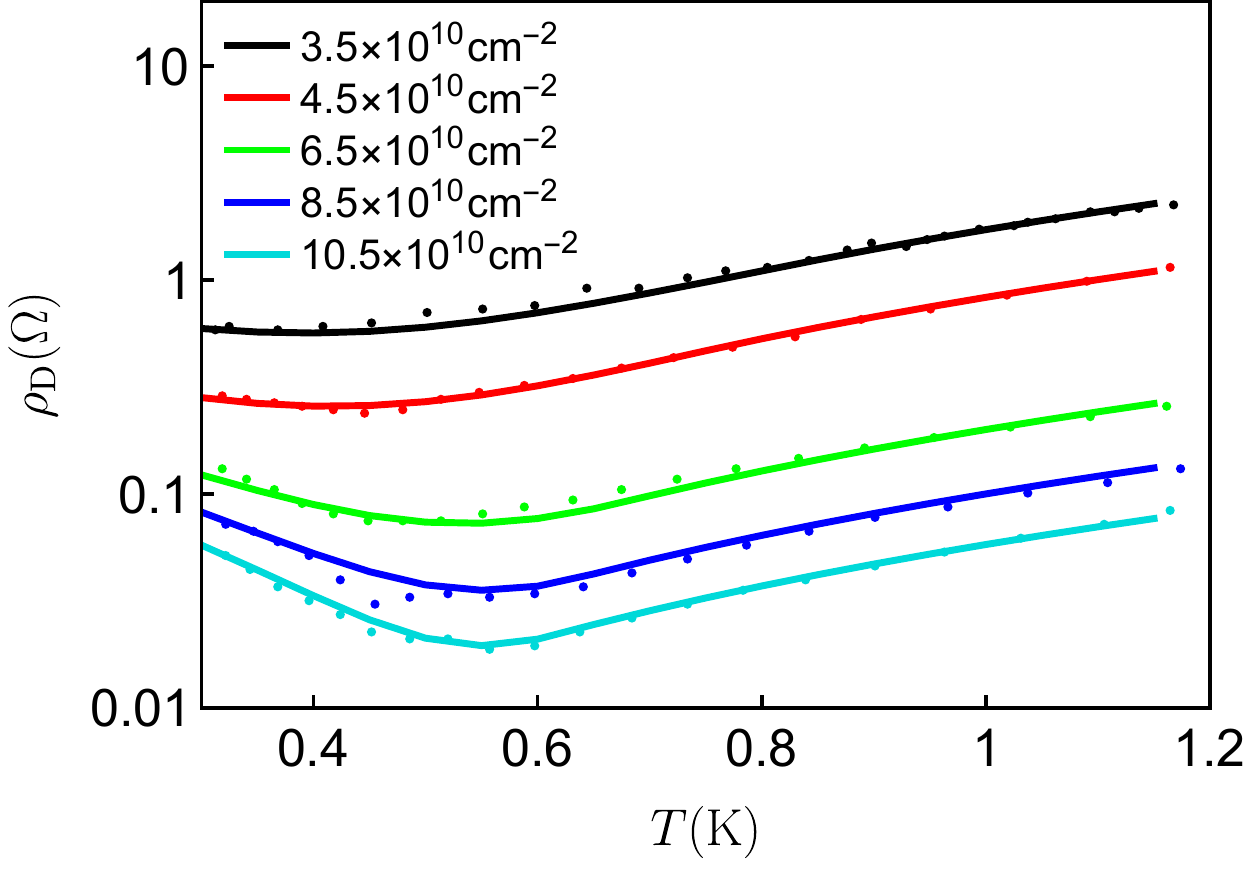}
\caption{\label{Imbalanced-density}
The temperature dependence of the drag resistivity for a series of density imbalances. The dotted lines are experimental data from Ref.~\cite{Christian-Lilly-PRB-2009}. The solid lines are fits to  our model with the same fitting parameters as in Fig.~\ref{Balanced-density}.
The fixed hole density is $n_\mathrm{h}=6.5\times 10^{10}\mathrm{cm}^{-2}.$ }
\end{center}
\vspace{-0.3in}
\end{figure}

\textcolor{blue}{\emph{Comparison  with experiment}}-- We compare our theory with semiconductor bilayer drag measurements,  
summarized in Figs.~\ref{Balanced-density} and \ref{Imbalanced-density}. The drag resistance $\rho_\mathrm{D}$
has a temperature dependence that clearly deviates from $T^2$ Fermi liquid behavior, and has a broad peak around zero density imbalance at the lowest temperature.
At high temperature, $\rho_\mathrm{D}$ approaches quadratic temperature dependence, in agreement with the picture of weakly coupled Fermi liquids. 
The anomalous low-temperature upturn is observed over a wide range of densities $n\approx 6\times 10^{10}$$\sim$$10^{11}\; \mathrm{cm}^{-2}$ 
and survives imbalances up to $\sim 30\%$.

To fit the experimental data we let 
$\rho_\mathrm{D}=\rho_\mathrm{D}^\mathrm{AR}+\rho_\mathrm{D}^\mathrm{C}$,
where $\rho_\mathrm{D}^\mathrm{A}$ is the contribution from AR and is given by Eq.~(\ref{RhoD}), while $\rho_\mathrm{D}^\mathrm{C}$ is due to Coulomb interactions and is assumed to have a quadratic temperature dependence: $\rho_\mathrm{D}^\mathrm{C}=z_\mathrm{C} T^2$. 
Here $z_\mathrm{C}$ is a fitting parameter which can be extracted from the high temperature $\rho_\mathrm{D}$ data. The masses of the charge carriers were set to $m_\mathrm{e}=m_\mathrm{h} =0.067\; m_0$, corresponding to electrons in $\hbox{GaAs}$. As a result, the temperature and doping dependence of $\rho_\mathrm{D}$ is determined by $\xi$ and $w$, as well as $\Delta_0(T)$ and $\delta_0(T)$. According to Eq.~(\ref{DeltaModule}), the latter two describe the dependence of $\Delta_\mathrm{A}$ on temperature and the local e-h imbalance. Their microscopic evaluation that takes into account the short-range disorder, long-range density variations, and screening, which is strongly affected by the presence of e-h condensate, is beyond the state-of-art microscopic approaches. For simplicity's sake we chose $\Delta_0(T)=3.06 \sqrt{T_0(T_0-T)}$ and $\delta_0(T)=1.2 \Delta_0(T)$ motivated by microscopic models in the BCS limit~\cite{Michael-Tinkham,Kinnunen_2018}. Here $T_0$ can be interpreted as the transition temperature. 

The parameters $w$ and $\xi$ can be estimated from the experimental setup. The percolation transition to the truly granulated insulating state is observed~\cite{Single-QW-Lilly,PhysRevLett.94.136401,PhysRevLett.100.016805} for electrons in QWs at $n_\mathrm{e}\approx 0.3\times 10^{10}\; \hbox{cm}^{-2}$ and suggests $w\approx0.2\sim0.7~\hbox{meV}$. Density variations are induced by charge inhomogeneities in $n^+$ GaAs cap layer which provides the electron doping to the QW. The cap layer is at distance $200~\hbox{nm}$ to the QW that suggests $\xi\approx100\sim 200~\hbox{nm}$. The cornerstone condition of our theory, $\delta\ll w\ll \epsilon_\mathrm{F}$, is very well satisfied~\footnote{It should be mentioned that the cornerstone condition $\delta\ll w\ll \epsilon_\mathrm{F}$ is also reasonably well satisfied in graphene/QW bilayers. Really, the anomalous Coulomb drag effect has been reported below $T\sim 5\;\hbox{K}$ and within the wide range of densities for Dirac holes around $n_\mathrm{h}=6.7\; 10^{11} \; \hbox{cm}^{-2}$. As a result, we could estimate $\delta_0\sim 0.4~\hbox{meV}$ and $\epsilon_\mathrm{F}\sim 3.7\; \hbox{meV}$. The density variations with height $w\sim 1\; \hbox{meV}$ are commonly present in graphene due to charged impurities in a substrate.. We keep these estimations as a benchmark and treat $w$ and $\xi$ as fitting parameters.}   

The contribution of AR to the drag resistivity $\rho_\mathrm{D}^\mathrm{A}$ depends smoothly  on $\xi$ and $w$,
which cannot be uniquely determined by fitting the temperature dependence $\rho_\mathrm{D}$ data for a given $n_\mathrm{e}$ and $n_\mathrm{h}$.  Instead, these parameters have been chosen~\footnote{We have assumed that $\xi$ and $w$ are doping and temperature independent. We neglect screening effects that self-consistently determine the density variations.} to achieve the best overall fit for the dependence of 
$\rho_\mathrm{D}(T)$ over a wide range of densities and density imbalances, with the result that $\xi=35\;\mathrm{nm}$ and $w=1.9\;\mathrm{meV}$ The final fit parameter $T_0$ is easily estimated from the position of the 
minimum in the temperature dependence of $\rho_\mathrm{D}$ and is adjusted separately for each curve.  
The temperature dependence of $\rho_\mathrm{D}$, presented in Fig.~\ref{Balanced-density} for the balanced case 
and in Fig.~\ref{Imbalanced-density} for the imbalanced case, is accurately captured by the theoretical model. The discrepancy between theory and experiment grows with decreasing e and h densities and can be explained by the proximity to the strong coupling regime, where the applicability of our phenomenological theory is limited. 

\textcolor{blue}{\emph{Discussion}--} The values $w=1.9\;\mathrm{meV}$ and $\xi=35\;\mathrm{nm}$ obtained from the fit are a bit different from the ones estimated from experiments and are at the edge of the applicability of our phenomenological theory. The discrepancy can be due to the oversimplification of $\Delta_0(T)$ and $\delta_0(T)$ as well as due to the large mismatch between the masses for e and h in semiconductor QWs that are not taken into account in our phenomenological theory.

Andreev-like reflection processes are accompanied by $\mathrm{e}$-$\mathrm{h}$ pair creation or annihilation processes that restore separate conservation of 
particle number in the electron and hole layers.  These processes play an essential role in the enhanced interlayer tunneling 
currents that are associated with enhanced tunneling between layers in bilayer electron-hole condensates~\cite{Tutuc-PRL-2018-E-H,Dmitry-PRB-2020}. In the drag geometry, however, the average tunneling current between layers is zero in the transport steady state.  The electric fields in the drag and drive layers drive steady state deviations from equilibrium in Bloch state occupation probabilities that are odd under inversion of momentum in both layers and there is not net generation of $e$-$h$ pairs.

To summarize, we have argued that the presence of density variations that are common in semiconductor QWs and in monolayer materials drives bilayer electron-hole condensates into a state with granulated order.  
In this phase, superfluidity is not supported, but the Coulomb drag effect is still strongly enhanced by AR at condensate grains. This scenario naturally explains the observed anomalous dependence of the drag resistivity on temperature and electron-hole imbalance. 

\textcolor{blue}{\emph{Acknowledgments}--}
We acknowledge fruitful discussions
with Alex Hamilton and support from
the Australian Research Council Centre of Excellence in
Future Low-Energy Electronics Technologies. AHM was supported by the U.S. Department of Energy under grant $\hbox{DOE-FG}02\hbox{-}02\hbox{ER}45958$, by Welch Foundation grant $\hbox{F-}1473$, and by Army Research Office (ARO) Grant No. $\hbox{W}911\hbox{NF-}17\hbox{-}1\hbox{-}0312$ (MURI). 

\bibliographystyle{apsrev4-1_our_style}
\bibliography{refs_AR}

\newpage

\begin{appendix}

\begin{widetext}
\section{Correlation function $F_\mathrm{\Delta}(\vec{R})$}
The Appendix proves that in the granulated phase the correlation functions for e-h condensate $\Delta(\vec{r})$ and auxiliary amplitude filed $ \langle\Delta_\mathrm{A}(\vec{r})\rangle$ are connected with $F_\mathrm{\Delta}(\vec{R})=\langle \Delta_\mathrm{A}(\vec{R}) \Delta_\mathrm{A}(0) \rangle- \langle \Delta_\mathrm{A}(0) \rangle^2$. Within the picture of well separated grains these fields $\Delta(\vec{r})$ and $ \Delta_\mathrm{A}(\vec{r})\rangle$ can be presented as follow 
\begin{equation}
\Delta(\vec{r})=\sum_i\Delta_\mathrm{g}(\vec{r}-\vec{r}_i)e^{i\phi_i}, \quad\quad \quad \quad  \Delta_\mathrm{A}(\vec{r})=\sum_i\Delta_\mathrm{g}(\vec{r}-\vec{r}_i).
\end{equation}
Here $\vec{r}_i$ and $\phi_i$ are position and phase of the order parameter for $i$-th grain. We assume that phases for different grains are uncorrelated while their profile $\Delta_\mathrm{g}(\vec{r})$ is the same. It should be noted that the \emph{auxiliary field} $\Delta_\mathrm{A}(\vec{r})$ is not equal to the amplitude of the order parameter as $\Delta_\mathrm{A}\neq |\Delta(\vec{r})|$. After averaging over positions and phases for rains we get the average values of the fields and their correlation functions 
\begin{equation}
\label{ACorrelationFunctions}
\ba
   &\dps  \langle\Delta(\vec{0}) \rangle= 0,  \quad \quad \quad \quad \quad \quad \quad \quad \;\; \langle \Delta_\mathrm{A}(\vec{0})\rangle=nF_1,  \\[3ex]
    &\dps \langle \Delta (\vec{R})\Delta^*(0)\rangle=nF_2(\vec{R}), \quad \quad \langle \Delta_\mathrm{A}(\vec{R})\Delta_\mathrm{A}(\vec{0})\rangle=n^2F^2_1+nF_2(\vec{R}).
    \ea
\end{equation}
Here the functions $F_1$ and $F_2(\vec{R})$ are given by
\begin{equation}
F_1=\int d\vec{r}\Delta_\mathrm{g}(\vec{r}), \quad \quad \quad \quad F_2(\vec{R})=\int d\vec{r}\Delta_\mathrm{g}(\vec{r})\Delta_\mathrm{g}(\vec{r}+\vec{R}).
\end{equation}
Explicit form of the correlation functions in (\ref{ACorrelationFunctions}) demonstrates that 
\begin{equation}
\label{CorrelatorIdentity}
    F_\Delta(\vec{R})=\langle \Delta(\vec{R})\Delta^*(0)\rangle=\langle \Delta_\mathrm{A}(\vec{R}) \Delta_\mathrm{A}(0)\rangle- \langle \Delta_\mathrm{A}(0) \rangle^2.
\end{equation}
It is noticed that the grain profile  $\Delta_\mathrm{g}(\vec{r})$ has not been specified. The relation Eq.~ (\ref{CorrelatorIdentity}) is only based on the assumption that each grain maintains the phase coherence and phases for different grains are uncorrelated.   
\section{Grain model with the Gaussian profile}
This Appendix presents the grain model with the Gaussian profile $\Delta_\mathrm{g}({\bm r})=\bar{\Delta}_\mathrm{g} e^{-r^2/a^2}$.   Here $\bar{\Delta}_\mathrm{g}$ is an amplitude of the order parameter at grain center and $a$ is its size.  The model is motivated by the Gaussian tail of the correlation function  $F_\mathrm{\Delta}(\vec{R})$ microscopically calculated in App. C, but not directly related with Eq.~(4) from the main part of the paper. The resulting functions $F_1$ and $F_2(\vec{R})$ are given by  
\begin{equation}
F_1=\int d\vec{r}\Delta_\mathrm{g}(\vec{r}) = \pi a^2 \bar{\Delta}_\mathrm{g}, \quad \quad \quad \quad F_2(\vec{R})=\int d\vec{r}\Delta_\mathrm{g}(\vec{r})\Delta_\mathrm{g}(\vec{r}+\vec{R})=\frac{1}{2}\pi \bar{\Delta}^2_\mathrm{g} a^2e^{-\frac{1}{2}\frac{R^2}{a^2}}. 
\end{equation}
As a result, the average amplitude of the the field $\Delta_\mathrm{A}$ and the correlation function $F_\mathrm{\Delta}(\vec{R})$ are given by
\begin{equation}
\langle\Delta_\mathrm{A}\rangle=\pi n a^2 \bar{\Delta}_\mathrm{g}, \quad\quad \quad F_\mathrm{\Delta}(\vec{R})=\frac{\pi n a^2}{2} \bar{\Delta}^2_\mathrm{g} e^{-\frac{1}{2}\frac{R^2}{a^2}}.
\end{equation}
The information about parameters for the grain model can be extracted from $\Delta_\mathrm{A}$ and $F_\mathrm{\Delta}(\vec{R})$. In particular, the developed phenomenological theory assumes the smallness for the diluteness parameter for grains, $n_\mathrm{g} a^2\ll1$, and the later can be extracted as $n_\mathrm{g} a^2= \langle \Delta_\mathrm{A}\rangle^2/2\pi\langle\langle \Delta^2_\mathrm{A}(0)\rangle\rangle$. This relation is not universal and is sensitive to details of the grain profile near its center. As a result, it will be used in App. C only for guiding and estimations. 

Within the model of grains with Gaussian profile the transport of the electron-hole bilayer can be analyzed.  We also incorporate the electron-hole mass asymmetry in Eqs.~(9-11) from the main part of the paper. If electron layer is the active layer, the AR transport scattering time is given by
\begin{equation}
\frac{1}{\tau_\mathrm{tr}^\mathrm{e}}=\frac{\pi m_\text{e}n_i}{\hbar^3}\bar{\Delta}^2_ga^4\mathrm{e}^{-(\bar{n}_\mathrm{e}+\bar{n}_\mathrm{h})}\mathrm{I}_1[2\sqrt{\bar{n}_\mathrm{e}\bar{n}_\mathrm{h}}].
\end{equation}
It vanishes for the point-like grains since in this case the probability of the Andreev reflection $P_\vec{q}$ becomes momentum independent and there no momentum transfer between the layers. As a result, the Andreev reflection mechanism contributes to the drag resistivity as follows
\begin{equation}
\label{AppDrag}
\rho^\mathrm{A}_\mathrm{D}=\frac{\hbar}{e^2}\frac{\pi^2 n_\mathrm{g} a^2 \bar{\Delta}^2_g}{4 \epsilon^\text{h}_\mathrm{g}\epsilon^\text{e}_\mathrm{g}}C(\bar{n}_\mathrm{e} ,\bar{n}_\mathrm{h}), \quad \quad C(\bar{n}_\mathrm{e} ,\bar{n}_\mathrm{h})= \frac{e^{-(\bar{n}_\mathrm{e}+\bar{n}_\mathrm{h})}}{\sqrt{\bar{n}_\mathrm{e} \bar{n}_\mathrm{h}} }\text{I}_1[2\sqrt{\bar{n}_\mathrm{e} \bar{n}_\mathrm{h}}],
\end{equation}
where $n_\mathrm{g}$ is a density for grains and  $\epsilon^\text{h}_\mathrm{g}=\hbar^2/2m_\text{h}a^2$ and $\epsilon^\text{e}_\mathrm{g}=\hbar^2/2m_\text{e}a^2$ are the corresponding energy scales; $\bar{n}_\mathrm{e}=\pi a^2 n_\mathrm{e}$, $\bar{n}_\mathrm{h}=\pi a^2 n_\mathrm{h}$ are dimensionless charge carrier density. As it is clearly seen, the mass of charge carriers only determines the prefactor in Eq.~(\ref{AppDrag}) that makes the electron-hole mass asymmetry neglected in the main part of the paper to be off importance. The dependence of the drag resistivity on electron and hole densities is determined by the factor $C(\bar{n}_\mathrm{e},\bar{n}_\mathrm{h})$ and $\text{I}_1(x)$ is the modified Bessel function of the first kind. The dependence of $C(\bar{n}_\mathrm{e},\bar{n}_\mathrm{h})$ on density is presented in Fig.~\ref{DensityGrains}. As a result, for given parameters of grains (that corresponds to their frozen distribution) $\rho_\mathrm{D}$ is temperature independent and very weakly depends on the density. Thus the anomalous dependence of the drag resistivity originates from the \emph{adjustment} of grains with temperature and e-h imbalance which is described by the correlation function of the order parameter given by Eq.~(8) from the main text and calculated in App.~C.       
\begin{figure}[h]
\begin{center}
\includegraphics[width=8cm]{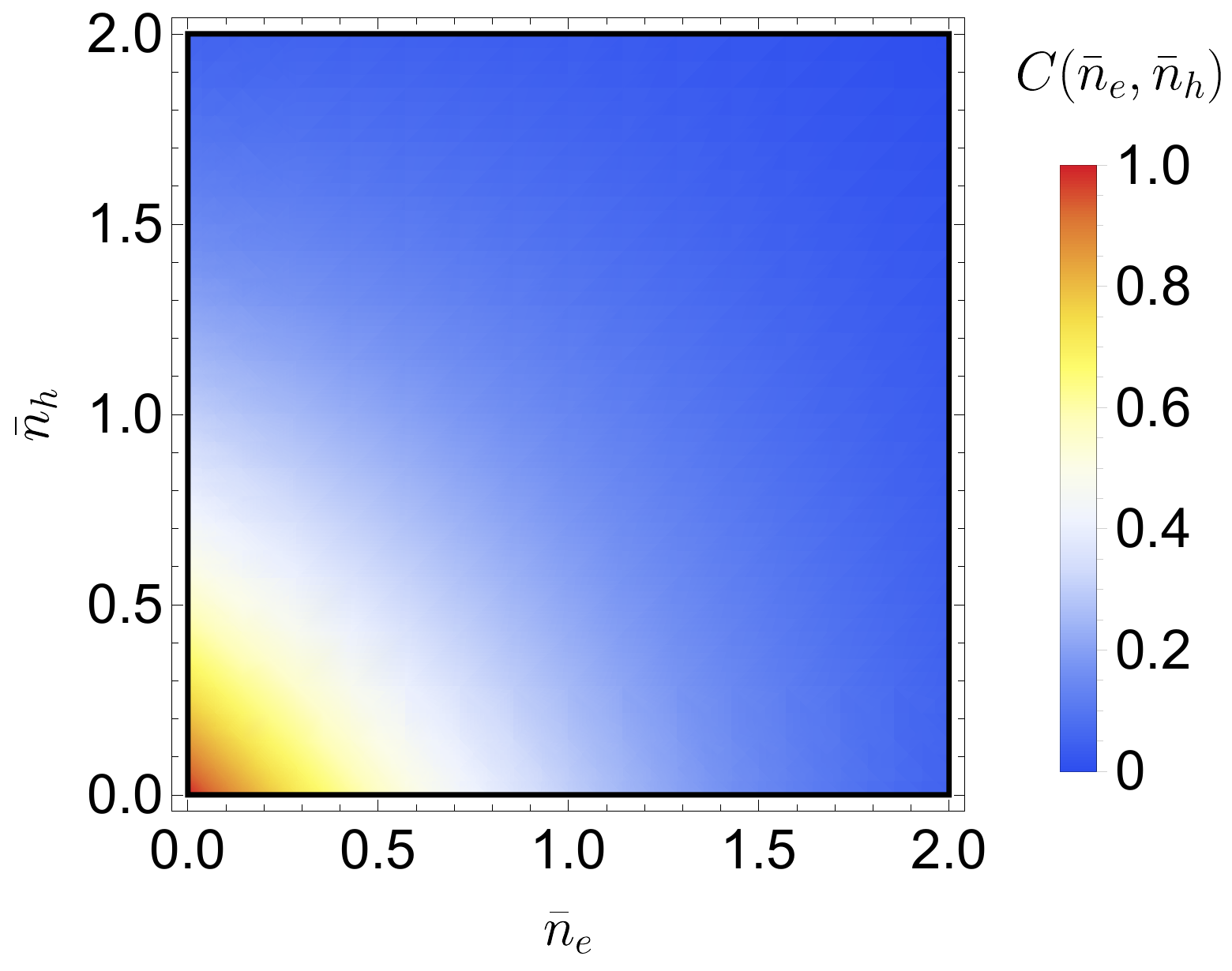}
\caption{\label{DensityGrains} The dependence of $C(\bar{n}_e,\bar{n}_h)$ on the density of two-layer system.}
\end{center}
\end{figure}
\section{The correlation function and effective field theory}
This Appendix presents a microscopic derivation of the correlation function for the order parameter $F_\Delta(\vec{R})=\langle \Delta(\vec{R})\Delta^*(0)\rangle$ . The order parameter is complex, but within the picture of well separated grains which maintain the local phase coherence, the phase is irrelevant. The role of the scattering potential is played solely by the position dependent absolute value of the order parameter $\Delta_\mathrm{A}(\vec{r})$. As it is argued in App.~A, the correlation function can be presented as follow $F_\Delta(\vec{R})=\langle \Delta_\mathrm{A}(\vec{R}) \Delta_\mathrm{A}(0)\rangle- \langle \Delta_\mathrm{A}^2(0) \rangle$. The amplitude of the order parameter $\Delta_\mathrm{A}$ is assumed to adjust to the local imbalance $\delta(\vec{r})$
\begin{equation}
\label{ApDeltaModule}
\Delta_\mathrm{A}({\bm r})=\Delta_0 \mathrm{exp}[-\delta^2/\delta^2_0].
\end{equation} 
Here $\Delta_0$ is temperature-dependent order parameter value in the absence of the mismatch. The averaging of the the correlation function is over the distribution for the local imbalance $\delta(\vec{r})=\delta_\mathrm{F}+\delta_\mathrm{V}(\vec{r})$, where $\delta_\mathrm{F}=\epsilon_\mathrm{F}^\mathrm{e}-\epsilon_\mathrm{F}^\mathrm{h}$ is responsible for the imbalance in average and $\delta_\mathrm{V}(\vec{r})=V_\mathrm{e}(\vec{r})-V_\mathrm{h}(\vec{r})$ for its variations. The latter is a random field with the Gaussian correlation function $\langle V(\vec{r}) V(0)\rangle= G_0(\vec{r})$ given by $G_0(\vec{r})=w^2\text{exp}\big[-r^2/\xi^2\big]$. As a result, the correlation function $\langle \Delta_\mathrm{A}(0)\Delta_\mathrm{A}(\vec{R})\rangle$ can be presented as
\begin{equation}
\langle \Delta_\mathrm{A}(0)\Delta_\mathrm{A}(\vec{R})\rangle  = \Delta_0^2\frac{\int D \delta_\mathrm{V}(\vec{r}) e^{-\frac{(\delta_\mathrm{F}+\delta_\mathrm{V}(0))^2}{\delta_0^2}-\frac{(\delta_\mathrm{F}+\delta_\mathrm{V}(\vec{R}))^2}{\delta_0^2}}\cdot e^{-S_0}}{\int D \delta_\mathrm{V}(\vec{r}) \cdot e^{-S_0} }  =\Delta_0^2\frac{\int D \delta_\mathrm{V}(\vec{r}) \cdot e^{-S}}{\int D \delta_\mathrm{V}(\vec{r}) \cdot e^{-S_0} },  
\end{equation}
Here $S$ is the effective action for a classical field $\delta_\mathrm{V}(\vec{r})$ given by 
\begin{equation}
\label{ActionS}
S=\underbrace{\frac{1}{2}\int d\vec{r}\int d\vec{r}'\delta_\mathrm{V}(
\vec{r})G^{-1}_0(\vec{r}-\vec{r}')\delta_\mathrm{V}(\vec{r}')}_{S_0 ,\quad \hbox{kinetic energy}} + \underbrace{\int d\vec{r} P(\vec{r})\delta_\mathrm{V}^2(\vec{r})}_{S_\mathrm{P}, \; \hbox{potential energy}} + \underbrace{\int d\vec{r} 2 P(\vec{r}) \delta_\mathrm{F} \delta_\mathrm{V}(\vec{r})}_{S_\mathrm{S}, \; \hbox{the source term}}+\frac{2 \delta_\mathrm{F}^2}{\delta_0^2}.    
\end{equation}
Here $G^{-1}_0(\vec{r})$ is the inverse Green function, while the effective potential and source term represent the sum of two repulsive point-like centers described by $P(\vec{r})=(\delta(0)+\delta(\vec{R}))/\delta_0^2$. The effective action~(\ref{ActionS}) is quadratic and the corresponding Gaussian integral can be evaluated with the relation
\begin{equation}
\int D\hat{x}\mathrm{e}^{-\frac{1}{2}\hat{x}^{T}\hat{A}^{-1}\hat{x}-\hat{B}\hat{x}}=\sqrt{\text{Det}A}\mathrm{e}^{\frac{1}{2}\hat{B}^{T}\hat{A}\hat{B}}.
\end{equation}
The result can be presented as 
$\langle \Delta_\mathrm{A}(0)\Delta_\mathrm{A}(\vec{R})\rangle  = \Delta_0^2 e^{-\Omega(\vec{R})}$ where $\Omega(\vec{R})$ is the effective thermodynamic potential given by 
\begin{equation}
\label{Chi}
\ba
\Omega(\vec{R})&\dps =\frac{1}{2}\mathrm{Tr}\ln(1+\hat{P}\hat{G}_0 )]+2 \delta_\mathrm{F} \int_{\vec{r}\vec{r}'}P(\vec{r})G_{\vec{r},\vec{r}'}P(\vec{r}') -\frac{2 \delta_\mathrm{F}^2}{\delta_0^2} = \frac{1}{2}  \int_0^1 d\lambda \int_\vec{r}P(\vec{r}) G_\lambda(\vec{r},\vec{r}) + 2 \delta_\mathrm{F} \int _{\vec{r}\vec{r}'} P(\vec{r})G(\vec{r},\vec{r}')P(\vec{r}') \\[3ex]
&\dps  -\frac{2 \delta_\mathrm{F}^2}{\delta_0^2} = -\frac{2}{\delta_0^2}  \int_0^1 d\lambda  [G_\lambda(\vec{0},\vec{0})+G_\lambda(\vec{R},\vec{R})] + \frac{2\delta_\mathrm{F}}{\delta_0^2}[G(\vec{0},\vec{0}) +G(\vec{0},\vec{R})+G(\vec{R},\vec{0}) +G(\vec{R},\vec{R})]  -\frac{2 \delta_\mathrm{F}^2}{\delta_0^2}. 
\ea
\end{equation}
Here we introduced auxiliary coupling constant $\lambda$ as $P(\vec{r})\rightarrow \lambda P(\vec{r})$ and employed the \emph{coupling constant integration} trick. The latter allows to effective reorder and re-sum the perturbation series for $\ln(1+\hat{P} G_0)$ which has unwanted $1/n$ factor in front of $n$-th term. The resumed series is presented in terms of renormalized Green function $\hat{G}_\lambda$ (and $\hat{G}\equiv \hat{G}_{\lambda=1}$ ) which satisfy the Dyson equation
\begin{equation}
G(\vec{r_1,r_2})=G_0(\vec{r_1-r_2})-\int d\vec{r}G^{-1}_0(\vec{r}_1-\vec{r}') \lambda P(\vec{r}')G(\vec{r}', \vec{r}_2).  
\end{equation}
The evaluation of (\ref{Chi}) involves only $G(\vec{0},\vec{0})=G(\vec{R},\vec{R})$ and $G(\vec{R},\vec{0})=G(\vec{0},\vec{R})$. The latter satisfy the closed system of algebraic equations
 \begin{equation}
 \ba
 G(\vec{0},\vec{0})= G_0(\vec{0},\vec{0})-2\lambda \delta_0^{-2}G_0(\vec{0},\vec{0}) G(\vec{0},\vec{0})-2\lambda \delta_0^{-2}G_0(\vec{0},\vec{R}) G(\vec{R},\vec{0}),\\[3ex]
 G(\vec{R},\vec{0})= G_0(\vec{R},\vec{0})-2\lambda \delta_0^{-2}G_0(\vec{R},\vec{0})G(\vec{0},\vec{0})-2\lambda \delta_0^{-2}G_0(\vec{R},\vec{R})G(\vec{R},\vec{0}).
 \ea
 \end{equation}
which result in 
\begin{equation}
G(\vec{0},\vec{0})=\frac{G_0(\vec{0}) -2\lambda \delta_0^{-2}G^2_0(\vec{R})+2\lambda \delta_0^{-2}G^2_0(\vec{0})}{[1+2\lambda \delta_0^{-2}G_0(\vec{0})]^2-(2\lambda \delta_0^{-2})^2G^2_0(\vec{R})},  \quad \quad \quad G(\vec{R},\vec{0}) =\frac{G_0(\vec{R})}{[1+2\lambda \delta_0^{-2}G(\vec{0})]^2-(2\lambda \delta_0^{-2})^2G^2(\vec{R})}.
\end{equation}
As a result, the effective thermodynamic potential is given by
\begin{equation}
\Omega(\vec{R}) = -\frac{1}{2}\ln\left[\frac{(\delta_0+2w (1+E))(\delta_0+2w (1-E))}{\delta_0^2}\right]-\frac{2\delta^2_\mathrm{F}}{\delta_0^2+2w^2(1+E)},
 \end{equation}
Here $E=e^{-R^2/\xi^2}$.The correlation function $\langle\Delta(\vec{R})\Delta(\vec{0}\rangle)$ is given by 
\begin{equation}\label{long}
\ba
&\dps \langle\Delta_\mathrm{A}(0)\Delta_\mathrm{A}(\vec{R})\rangle=\Delta^2_0\frac{\delta_0}{\sqrt{(\delta_0+2w (1+E))(\delta_0+2w (1-E))}}\mathrm{e}^{-\frac{2\delta^2_\mathrm{F}}{\delta_0^2+2 w^2(1+E)}}.
\ea 
\end{equation}
In the limit of $R\rightarrow \infty$, the correlation function decouples $\langle \Delta(\vec{R})\Delta(\vec{0})\rangle\rightarrow \langle \Delta\rangle^2$. Therefore the averaged value $\langle \Delta_\mathrm{A}\rangle$ and the correlation are given by \begin{equation}\label{correlationSM}
\langle\Delta_\mathrm{A} \rangle = \Delta_0 \frac{ \delta_0 e^{-\frac{\delta^2_\mathrm{F}}{\delta^2_0+2w^2}} }{\sqrt{\delta^2_0+2w^2}},  \quad \quad\quad  F_\Delta(\vec{R})=\Delta^2_0\Big\{\frac{ \delta^2_0 e^{-\frac{2\delta_\mathrm{F}^2}{\delta^2_0+2w^2(1+E)}}}{\sqrt{\big[\delta^2_0+2w^2(1+E)\big]\big[\delta^2_0+2w^2(1-E)\big]}}-\frac{\delta^2_0 e^{-\frac{2\delta^2_F}{\delta^2_0+2w^2}}}{\delta^2_0+2 w^2}\Big\}.
\end{equation}
The function $F_\mathrm{\Delta}(\vec{R})$ is plotted and analyzed in the main text. It has the Gaussian tail at $R\ll \xi_\mathrm{\Delta}$ where $\xi_\mathrm{\Delta}=\sqrt{3} \delta_0 \xi/w$ can be interpreted as a character grain size. This suggests to use the relation for the dilutness parameter of grains $n_\mathrm{g} a^2= \langle \Delta_\mathrm{A}\rangle^2/2\pi\langle\langle \Delta^2_\mathrm{A}(0)\rangle\rangle\approx h_0/w$, which we have derived in App. B within the model of random grains. Relation can be dependent on the grain profile near its center and needs to be used only for guiding and estimations, however the smallness of $h_0/w$ justifies the picture of well separated grains for e-h condensate. 

\section{Dependence of drag resistivity $\rho_\mathrm{D}$ on e-h imbalance}
The Appendix presents dependence of the drag resistivity on temperature and electron-hole imbalance that is shown in Fig.~\ref{DesnityPlot}. The parameters in the calculations are  $T_0=1\mathrm{K}$, $\epsilon_{\mathrm{F}}=0.30~\mathrm{meV}$, $\delta_0=0.3\mathrm{meV}$, $w =3~\mathrm{meV}$, $\xi=60\;\mathrm{nm}$, average density for e and h  $\bar{n}=6\times10^{11}\mathrm{cm}^{-2}$ . The drag resistivity experiences an upturn in the vicinity of $T_0$ that is robust to the imbalance. In fact, locally e-h pairing is very sensitive to the imbalance and is strongly suppressed if the latter exceeds $\delta_0$. However, as it is clearly seen in Fig.~(\ref{DesnityPlot}), $\rho_\mathrm{D}$ survives imbalance until $\delta_\mathrm{F}\sim w$ with $w\gg \delta_0$. In this case $\delta_\mathrm{F}\gtrsim w$ the probability to find the spot with matching charge carrier density for e and h is exponentially small.

\begin{figure}
\begin{center}
\label{DesnityPlot}
\includegraphics[width=8cm]{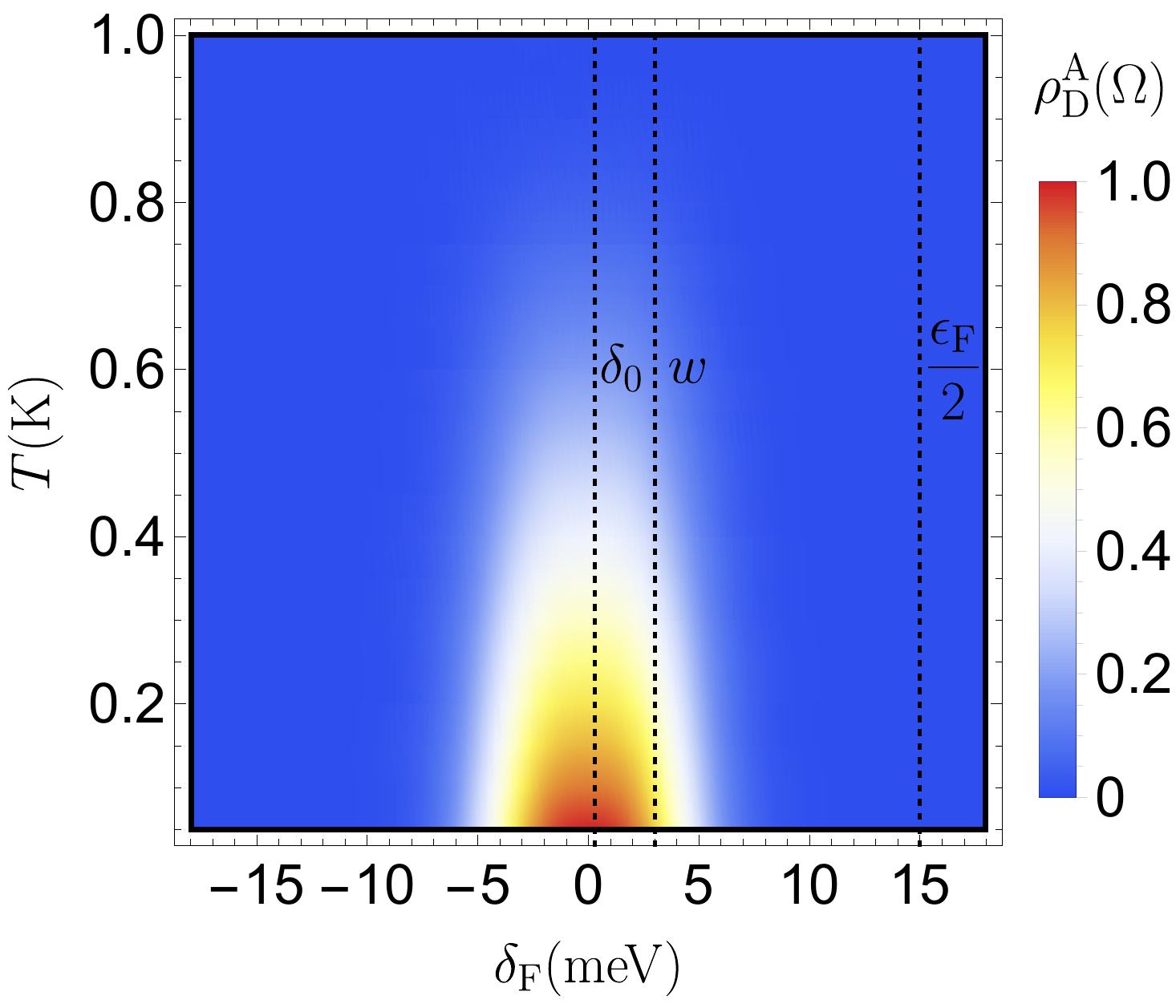}
\caption{\label{Density} The temperature $T$ and Fermi energy difference $\delta_\mathrm{F}$ dependence of drag resistivity induced by Andreev reflection of charge carriers at condensate grains. Parameters for calculations are $T_0=1\mathrm{K}$, $\epsilon_{\mathrm{F}}=30~\mathrm{meV}$,$w =3~\mathrm{meV}$, $\delta_0=0.3\mathrm{meV}$,  $\xi=60\;\mathrm{nm}$, average and density for e and h  $\bar{n}=6\times10^{11}\mathrm{cm}^{-2}$.}
\end{center}
\end{figure}

\end{widetext}
\end{appendix}

\end{document}